\documentclass[aps,prl,english,superscriptaddress,twocolumn,tightenlines,showpacs,floatfix,amsfonts]{revtex4-1}
\usepackage[T1]{fontenc}
\usepackage[latin9]{inputenc}
\usepackage[dvipsnames,svgnames]{xcolor}
\usepackage{float}
\usepackage{graphicx}
\usepackage{textcomp}
\usepackage{mathrsfs}
\usepackage{mathtools}
\usepackage{amsmath,amssymb,amsthm,amscd,bm}
\usepackage{txfonts} 
\usepackage[colorlinks,bookmarks=false,citecolor=blue,linkcolor=red,urlcolor=blue]{hyperref}

\newcommand{\changes}[1]{{\color{black}#1}}

\begin{document}
\title{Tangle of Spin Double Helices in the Honeycomb Kitaev-$\Gamma$ Model}
\author{Jheng-Wei Li}
\affiliation{Arnold Sommerfeld Center for Theoretical Physics, University of Munich, Theresienstr. 37, 80333 M\"{u}nchen, Germany}
\affiliation{Munich Center for Quantum Science and Technology (MCQST), Schellingstr. 4, 80799 M\"{u}nchen, Germany}

\author{Nihal Rao}
\affiliation{Arnold Sommerfeld Center for Theoretical Physics, University of Munich, Theresienstr. 37, 80333 M\"{u}nchen, Germany}
\affiliation{Munich Center for Quantum Science and Technology (MCQST), Schellingstr. 4, 80799 M\"{u}nchen, Germany}

\author{Jan von Delft}
\affiliation{Arnold Sommerfeld Center for Theoretical Physics, University of Munich, Theresienstr. 37, 80333 M\"{u}nchen, Germany}
\affiliation{Munich Center for Quantum Science and Technology (MCQST), Schellingstr. 4, 80799 M\"{u}nchen, Germany}

\author{Lode Pollet}
\affiliation{Arnold Sommerfeld Center for Theoretical Physics, University of Munich, Theresienstr. 37, 80333 M\"{u}nchen, Germany}
\affiliation{Munich Center for Quantum Science and Technology (MCQST), Schellingstr. 4, 80799 M\"{u}nchen, Germany}
\affiliation{Wilczek Quantum Center, School of Physics and Astronomy, Shanghai Jiao Tong University, Shanghai 200240, China}

\author{Ke Liu}
\email{ke.liu@lmu.de}
\affiliation{Arnold Sommerfeld Center for Theoretical Physics, University of Munich, Theresienstr. 37, 80333 M\"{u}nchen, Germany}
\affiliation{Munich Center for Quantum Science and Technology (MCQST), Schellingstr. 4, 80799 M\"{u}nchen, Germany}

\date{\today}
\begin{abstract}
We investigate the ground-state nature of the honeycomb Kitaev-$\Gamma$ model in the material-relevant parameter regime through a combination of analytics and classical and quantum simulations.
We find the classical model is imprinted with a tangle of highly structured spin double helices.
This helix tangle consists of $18$ inequivalent helices and features modulation of multiple rotation axes, a spontaneous anisotropy in spacial periodicities, and a ${\rm sgn}(\Gamma)$-determined chirality pattern.
Infinite PEPS simulations with clusters up to $36$ sites identify hallmarks of this unprecedented many-body order in the quantum spin-$1/2$ model.
Our findings provide a fresh perspective of the Kitaev-$\Gamma$ model and enrich the physics of Kitaev magnetism.
\end{abstract}
\maketitle

\paragraph{Introduction.}
The honeycomb Kitaev-$\Gamma$ Hamiltonian is a paradigmatic model for the physics of two-dimensional Kitaev magnets~\cite{Jackeli09, Chaloupka10, Rau14, Rau16, Wang17, Takagi19, Winter17b}.
The Kitaev exchange leads to an exactly solvable spin liquid~\cite{Kitaev06} and can be realized in $d$-electron transition-metal compounds with edge-shared geometry~\cite{Jackeli09}.
However, symmetries of real materials permit a generic existence of a Heisenberg and an off-diagonal $\Gamma$ term~\cite{Rau14, Rau16, Wang17} which may dramatically modify the desired spin-liquid ground state.
In particular, in the prime candidate material $\alpha\text{-RuCl}_3$~\cite{Banerjee16,Banerjee17,Zheng17,Kasahara18,Yokoi21,Bachus20,Czajka21,Czajka22,Bruin22,Lefrancois22}, the $\Gamma$ exchange is comparable with the Kitaev one~\cite{Ran17, Yadav16, Kim16, Winter16, Winter17, Maksimov20, Laurell20}.
Theoretical studies further suggest the competition between $\Gamma$ and Kitaev interactions can induce novel exotic phases beyond a single spin liquid~\cite{Maksimov20,Laurell20, Rusnacko19, Jiang19, Gordon19, Lee20, Gohlke20, Wang19, Buessen21, Wang20, Sorensen21, Chern20, Chern21b, Liu21, Rao21b, Lampen18, Rayyan21, Chen22}, while the Heisenberg term stabilizes regular magnetic orders~\cite{Chaloupka10}. 
Nonetheless, state-of-the-art numerical methods, including exact diagonalization~\cite{Lampen18, Rusnacko19}, tensor network~\cite{Gordon19,Jiang19,Lee20,Gohlke20}, variational Monte Carlo~\cite{Wang19}, and functional renormalization group~\cite{Buessen21} techniques, have yielded highly diverse results for the spin-$1/2$ Kitaev-$\Gamma$ model, leaving the quantum phase diagram obscure.
Consensus exists for a Kitaev spin liquid (KSL) at finite $\Gamma$, but even its extent is debated.

\begin{figure}[!t]
  \centering
  \includegraphics[width=0.50\textwidth]{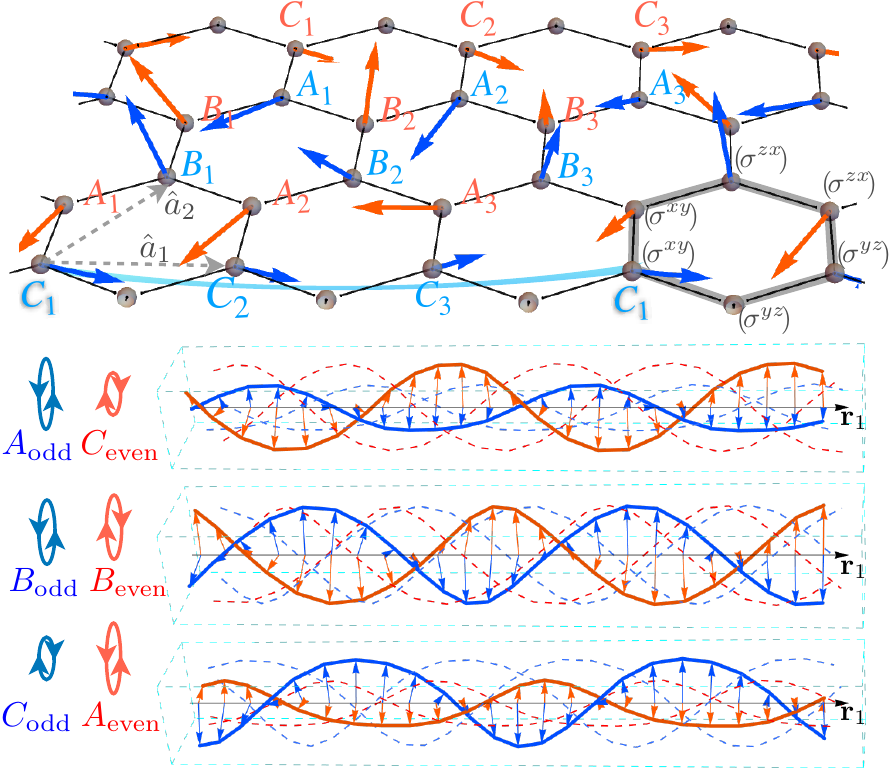}
  \caption{A classical ground state of the honeycomb Kitaev-$\Gamma$ model at $\Gamma=-K$ on a $L=72$ lattice.
  \changes{Upper panel: Structure of the $3\times 3$ supercell determined from the helical axes (longitudinal components) of spins.
  $A, B, C$ label three independent directions of the helical axes, and $j= 1,2,3$  distinguishes different orientations due to the hidden symmetry.
  Blue and red colors mark the odd and even honeycomb sublattices and the helices' chirality.
  A supercell specifies $18$ {\it inequivalent} helices as spins form a helix {\it only if} they belong to the same sublattice, e.g., the linked blue $C_1$ spins.
  The grey hexagon marks a unit cell of the $6$-site hidden symmetry transformation.
  Lower panel: Spiral (transverse) components of the $18$ helices in their respective sublattice basis.
  Each helix consists of $\frac{L}{3}$ spins due to the $3\times 3$  supercell structure.
  The helical pitches are {\it spontaneously anisotropic} and are $\frac{L}{6}$ and $\frac{L}{3}$ supercells in size along directions of the two lattice vectors $\hat{a}_{1,2}$.
  Cycles on the side reflect the strength and staggered chirality of the corresponding helix.
  The helix ensemble can be viewed as nine pairs of double helices.}}
  \label{fig:config}
\end{figure}

In this Letter, we conduct large-scale classical and quantum simulations to unravel the ground-state nature of the honeycomb Kitaev-$\Gamma$ model in the most puzzling yet material-relevant regimes.
We demonstrate that the classical ground state imprints a tangle of emergent \emph{spin double helices} (Fig.~\ref{fig:config}).
It exhibits several prominent characteristics, including modulation of multiple helical axes, anisotropic periodicities, and a ${\rm sgn}(\Gamma)$-determined chirality pattern,
\changes{as summarized in the ansatz Eq.~\eqref{eq:S_trans}.} 
The emergence of such a sophisticated helix tangle poses fundamental challenges to quantum algorithms and is the source of non-coherent observations in the literature.
Nevertheless, our infinite projected entangled pair states (iPEPS) calculations identify signatures of this unprecedented many-body order, showing that it can survive quantum fluctuations in the \mbox{spin-$1/2$} case.

\paragraph{Model and symmetry.}
The Kitaev-$\Gamma$ model on a honeycomb lattice comprises two bond-dependent terms: a directional Ising-type interaction and a symmetric off-diagonal $\Gamma$ interaction,
\begin{align}\label{eq:model}
	H = \sum_{\langle i j\rangle_{\gamma}} K S_{i}^{\gamma} S_{j}^{\gamma}+ \Gamma \left(S_{i}^{\alpha} S_{j}^{\beta}+S_{i}^{\beta} S_{j}^{\alpha}\right).
\end{align} 
Here, $\gamma$ labels the three different types of bonds, and $\alpha, \beta, \gamma \in \{x,y,z\}$ are mutually exclusive.
For example, the local Hamiltonian on a $z$-bond reads
$H_z = K S_{i}^{z} S_{j}^{z} + \Gamma \left(S_{i}^x S_{j}^y + S_{i}^y S_{j}^x \right)$.
For simplicity, we assume the coupling strengths on the three bonds are uniform.
Nevertheless, one expects that a small bond anisotropy (e.g., $K_z \neq K_{x, y}$) does not alter the nature of the underlying phases as in the pure Kitaev model~\cite{Kitaev06}.

The Hamiltonian Eq.~\eqref{eq:model} has a \emph{hidden global} symmetry that intertwines the spacial and spin spaces and leads to an additional two-fold degeneracy.
This is seen by rewriting Eq.~\eqref{eq:model} as 
$H = \sum_{\langle i j\rangle_{\gamma}} \mathbf{S}_i \cdot \hat{J}^{\gamma}_{ij} \mathbf{S}_j$, with three exchange matrices 
\begin{align}
\hat{J}^x_{ij} = \left[ 
\begin{smallmatrix}
K & &  \\
 &  & \Gamma \\
 & \Gamma  & 
\end{smallmatrix}
\right], \ 
\hat{J}^y_{ij} = \left[ 
\begin{smallmatrix}
 & & \Gamma \\
 & K &  \\
 \Gamma &  & 
\end{smallmatrix}
\right],  
\hat{J}^z_{ij} = \left[ 
\begin{smallmatrix}
  & \Gamma &  \\
\Gamma & &  \\
 &  & K
\end{smallmatrix}
\right].
\end{align}

One can verify that $\hat{J}^{\gamma}_{ij}$, hence the local environment of spins, is invariant under the transformations
\begin{gather}
\sigma^{\alpha\beta}_i \hat{J}^{\gamma}_{ij} \sigma^{\alpha\beta}_j = \hat{J}^{\gamma}_{ij}, \label{eq:symmetry1} \\ 
\sigma^{\alpha\gamma}_i \hat{J}^{\gamma}_{ij} \sigma^{\gamma\beta}_j = \hat{J}^{\gamma}_{ij}, \label{eq:symmetry2}
\end{gather}
where $\sigma^{\alpha\beta}_i$ denotes spin reflections transforming a spin as $\sigma^{\alpha\beta}_i S^\gamma_i = -S^\gamma_i$.
These transformations alternate the three spin reflections in accordance with the index rule of Eqs.~\eqref{eq:symmetry1} and \eqref{eq:symmetry2}, until covering the entire lattice (see the SM~\cite{SM} for an example).
In contrast to usual global symmetries, such as the time reversal and homogeneous spin rotations, this hidden symmetry can modify correlations in the system and make two \emph{distinct} orders degenerate.

\paragraph{Selected parameter region.}
The main open problems in constructing the phase diagram of the spin-$1/2$ Kitaev-$\Gamma$ model can be viewed from three fronts.
First, although mounting numerical evidence suggests an extended KSL regime under a small $\Gamma$, different algorithms find different extents of the regime~\cite{Gordon19, Wang19, Lee20, Gohlke20, Buessen21}.
Furthermore, in the large $\Gamma$ limits, the fate of classical $\Gamma$ spin liquids~\cite{Rousochatzakis17, Liu21, Saha19} is unsettled and prone to simulation techniques~\cite{Luo21,Discussion}.
Away from these two limits, a diversity of candidate ground states have been suggested, including non-Kitaev spin liquids, quantum paramagnets, incommensurate or spiral orders, and various magnetic states~\cite{Gordon19, Wang19, Lee20, Gohlke20, Buessen21, Jiang19}.
As these open problems cannot be resolved in a single work, we focus our efforts on the regime where both interactions are sizable and competing, which is of the highest relevance for real materials.
For convenience, we parameterize the two interactions using an angle parameter $\theta$ as $K = \sin\theta$, $\Gamma = \cos\theta$.
We consider the frustrated region with a ferromagnetic $K$ and an anti-ferromagnetic $\Gamma$ ($1.5\pi < \theta < 2\pi$) as realized in the $d$-electron ${\rm Ir}$- and ${\rm Ru}$-based compounds~\cite{Winter17, Yadav19, Laurell20}.
Through a mapping $\theta \rightarrow \theta + \pi$ and simultaneously $\mathbf{S}_{2i} \rightarrow \mathbf{S}_{2i}$, $\mathbf{S}_{2i+1} \rightarrow -\mathbf{S}_{2i+1}$~\cite{Mapping}, the results also enlighten the physics at  $0.5\pi < \theta < \pi$,
\changes{where $2i$ and $2i+1$ denote the even and odd honeycomb sublattices, respectively.}
Other parameter regions with $\theta \in (0, 0.5 \pi)$ and $(\pi, 1.5 \pi)$ are unfrustrated and  develop strong $120^{\circ}$-type magnetic orders~\cite{Rusnacko19, Liu21}.

\begin{figure}[t]
  \centering
  \includegraphics[width=0.45\textwidth]{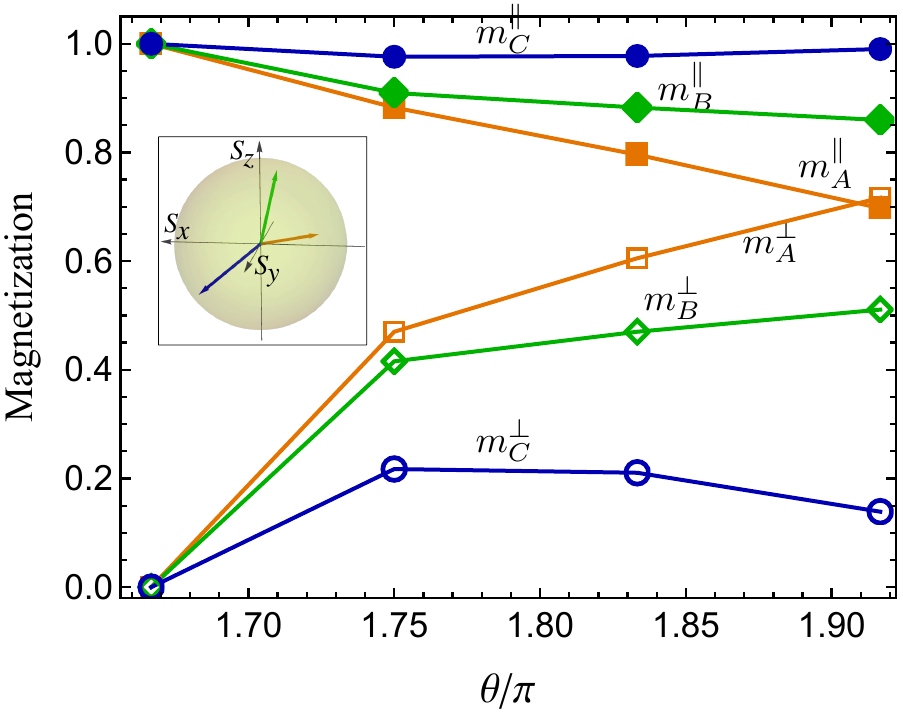}
  \caption{Magnetizations of the classical spin-helix tangle with comparable Kitaev and $\Gamma$ interaction, $K = \sin\theta, \Gamma = \cos\theta$. The longitudinal and spiral components the order are measured by $m^{\parallel}_{A,B,C}$ and $m^{\perp}_{A,B,C}$, respectively, leading to saturated total magnetizations $\big(m^{\parallel}_{A,B,C}\big)^2 + \big(m^{\perp}_{A,B,C}\big)^2 = 1$. The spiral component at $\theta \approx 1.67 \pi$ is small but non-vanishing (see Fig.~\ref{fig:osc}). The inset exemplifies the orientations of $\mathbf{m}^{\parallel}_{A}$, $\mathbf{m}^{\parallel}_{B}$ and $\mathbf{m}^{\parallel}_{C}$ (distinguished by colors) at $\theta = 1.75 \pi$.}
  \label{fig:mags}
\end{figure}

\paragraph{Tangle of spin helices.}
We first discuss the classical ground states in the selected parameter regime, \changes{which is crucial for understanding the quantum ground states.}
The growing interest in the classical Kitaev-$\Gamma$ model is also rewarded with rich physics~\cite{Chern20, Liu21, Chern21b, Rao21b, Lampen18, Rayyan21, Chen22}.
In particular, two recent works~\cite{Lampen18,Chern20} based on analysis of small systems reported various large-unit-cell states, including two degenerate $6$- and $18$-site structures.
By examining large systems at temperature $T = 10^{-3}\sqrt{K^2 + \Gamma^2}$, these two states were further mapped to a frustrated phase spanning over $1.58\pi \lesssim \theta < 2\pi$~\cite{Liu21}.
This phase can be understood by the competition between two classical spin liquids, and its magnetization exhibits an intrinsic undersaturation indicating the lack of perfect translationally invariant order~\cite{Liu21}.

We now reveal the missing magnitude encodes \changes{the essential nature} of the classical ground state, which only manifests at very large lattices and cold temperatures.
We utilize parallel tempering Monte Carlo methods to reach temperatures down to $T = 10^{-5}\sqrt{K^2 + \Gamma^2}$ for systems of linear size up to \changes{$L = 72, \, 90, \, 108$}, under periodic boundary conditions.
We further cool the system to $T \rightarrow 0$ by eliminating remaining thermal noise; see the SM~\cite{SM} for details of simulations.

To understand the structure of the ordering, we first discuss its longitudinal components, which lead to a $3\times 3$ supercell containing $18$ sublattices as depicted in Fig.~\ref{fig:config}.
The corresponding magnetic moments $\mathbf{m}^{\parallel}_\mu$ are obtained by averaging spins over all supercells as
$\mathbf{m}^{\parallel}_{\mu} = \frac{1}{N_{\rm cell}} \sum_{\rm cells} \mathbf{S}_\mu$,
where $\mu$ distinguishes the $18$ sublattices.
These magnetic moments are the source of the stable $\frac{2}{3}\mathbf{M}$ magnetic Bragg peaks reported in the literature~\cite{Liu21, Chern20}, and also act as the rotation axes of the highly structured spin helices.

There are two degenerate magnetization patterns due to the hidden symmetry.
The example in the upper panel of Fig.~\ref{fig:config} shows a relatively simpler pattern whose helical axes fall into three different orientations, labeled with italic letters $A, B, C$.
In this case, all $A_j$ spins have identical longitudinal components
$\mathbf{S}^{\parallel}_{A_j} = \mathbf{m}^{\parallel}_{A_j}$;
similarly for $B_j$ and $C_j$ spins.
Its degenerate state is obtained by applying the transformations Eqs.~\eqref{eq:symmetry1} and~\eqref{eq:symmetry2} as indicated by the grey hexagon, leading to nine distinct rotation axes distinguished by the numerical subscripts $j = 1, 2, 3$.
Since the hidden symmetry is exact, one can always infer all helical axes from three independent moments $\mathbf{m}^{\parallel}_{A,\, B, \, C}$. 
These magnetic moments are non-collinear and non-coplanar and vary with $K/\Gamma$ in both magnitude and orientation.
Fig.~\ref{fig:mags} measures their magnitudes $m^{\parallel}_{A,\, B, \, C}$ in the helical phase and shows an explicit example of their orientations.
A procedure to semi-analytically estimate their directions is provided in the SM~\cite{SM}.

It is important to note that the supercell is only defined for the longitudinal order.
The full ordering does not have a translational symmetry due to its incommensurate transverse components, as illustrated in the lower panel of Fig.~\ref{fig:config}.
Moreover, the $18$ helices specified by $18$ sublattices are {\it inequivalent} despite that their rotation axes may have the same orientations.
Namely, there is a $3\times 3$ modulation in the helical axes, and spins form a helix {\it only if} they belong to the same sublattice, instead of with their neighbors.

\begin{figure}[t]
  \centering
  \includegraphics[width=0.45\textwidth]{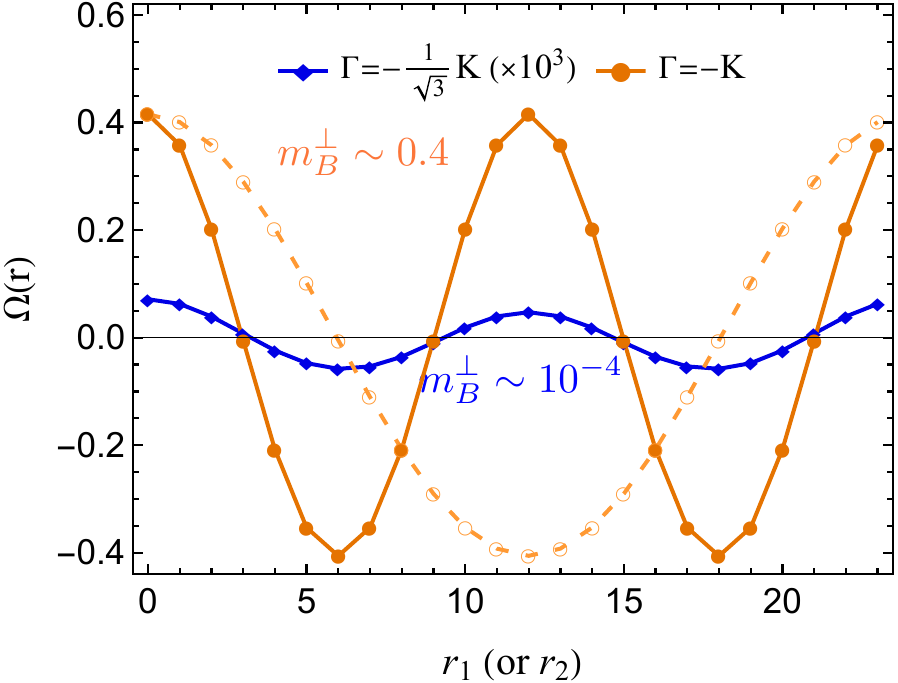}
  \caption{The \changes{spiral correlation function} $\Omega_\mu(\mathbf{r})$ for B-helices at $\Gamma = -K/\sqrt{3}$ (blue; rescaled by $10^3$) and $-K$ (orange) ($\theta = 1.67\pi, 1.75\pi$). The oscillation and the amplitude reflect the periodicity of the helices and strength of the spiral magnetization $m^{\perp}_{B}$, respectively. Filled and open symbols are measured along spontaneously chosen lattice $\mathbf{r}_1$ and $\mathbf{r}_2$ directions, \changes{with pitch sizes of $\frac{L}{6}$ and $\frac{L}{3}$ supercells}, respectively. A $L = 72$ lattice is considered, and $r_1, r_2 = 0, 1, \dots, \frac{L}{3}-1$.}
  \label{fig:osc}
\end{figure}

To describe the spiral components, it is convenient to work in sublattice coordinates where $\mathbf{m}^{\parallel}_{\mu}$ is rotated to
$\widetilde{\mathbf{m}}^\parallel_\mu = (0, \ 0, \ m^\parallel_{\mu})$ for each sublattice.
Then the spiral component of a spin can be formulated by an ansatz
\begin{align}\label{eq:S_trans}
	\widetilde{\mathbf S}^{\perp}_{\mu,\mathbf{r}} = m^{\perp}_\mu 
	\left(\cos\tfrac{6\pi}{L}(2r_1 + r_2),
	 {\rm sgn}(\Gamma)^\eta \sin\tfrac{6\pi}{L}(2r_1 + r_2), 0
	\right).
\end{align}
Here, $m^{\perp}_\mu = \frac{1}{N_{\rm cell}} \sum_{\rm cells} |\widetilde{\mathbf S}^{\perp}_{\mu}|$ defines a spiral magnetization, which fulfils
$\big(m^{\parallel}_\mu\big)^2 + \big(m^{\perp}_\mu\big)^2 = 1$
as measured in Fig.~\ref{fig:mags};   
$\mathbf{r} = (r_1, r_2)$ with $r_1, r_2 = 0, 1, \dots, \frac{L}{3}-1$
labels the $3\times 3$ supercells, 
and $\eta = 0, 1$ distinguishes the even and odd honeycomb sublattices.

One notices that the relative chirality of the spin helices is determined by the sign of $\Gamma$.
Hence, in the parameter region $\Gamma > 0$ and $K < 0$, spin helices living on the odd and even honeycomb sublattices (blue and red sites in Fig.~\ref{fig:config}) can be viewed as nine pairs of double helices with opposite chirality.
We also verify that the other frustrated region with $\Gamma < 0$ and $K > 0$ has a uniform chirality pattern.

Another highly unusual feature of the helical phase is that it displays a \emph{spontaneously} anisotropy in spatial periodicity.
The helical pitch sizes are $\frac{L}{6}$ and $\frac{L}{3}$ supercells along the directions of the two lattice vectors, respectively.
Namely, a helix completes two periods along a spontaneously chosen $\mathbf{r}_1$ direction but only one period along the other direction, despite the same amplitude.

Such a periodicity anisotropy has been encoded in the ansatz Eq.~\eqref{eq:S_trans} and can be shown explicitly by measuring a \changes{spiral correlation function}
\begin{align}
	\Omega_{\mu} (\mathbf{r}) = \frac{1}{m^{\perp}_\mu} \left( \mathbf{S}_{\mu, \mathbf{r}_0} \cdot \mathbf{S}_{\mu, \mathbf{r}_0 + \mathbf{r}} - |\mathbf{m}^{\parallel}_\mu|^2\right),
\end{align}
\changes{where the distance $\mathbf{r}$ is measured in unit of supercells.
$\Omega_{\mu} (\mathbf{r})$ shall develop a cosine curve reflecting the periodicity of the underlying spin helix,}  
and its amplitude indicates strength of the spiral magnetization $m^{\perp}_\mu$.
In Fig.~\ref{fig:osc}, we compare the behaviors of $\Omega_{\mu} (\mathbf{r})$ for two $B$-helices along spontaneously chosen $\mathbf{r}_1$ and $\mathbf{r}_2$ at $\Gamma = -K$ (orange curves) for a $L=72$ lattice, by which the spontaneous anisotropy is confirmed.

We stress that the helical order is {\it genuinely incommensurate}. Thus, in any finite-size simulations the observed pitch sizes $\frac{L}{6}$ and $\frac{L}{3}$ will grow linearly with $L$; see the SM~\cite{SM}.

In Fig.~\ref{fig:osc}, we also measure $\Omega_{\mu} (\mathbf{r})$ for a $B$-helix at $\Gamma = -K/\sqrt{3}$ (blue curve).
Despite an extremely small $m^{\perp}_B \sim 10^{-4}$, the expected cosine oscillation remains developed.
This means that the helicity is an imprinted feature of the entire frustrated phase $1.58\pi \lesssim \theta < 2\pi$, while it may be easily overlooked at weaker $\Gamma$ values.

\begin{figure}[t]
  \centering
  \includegraphics[width=0.47\textwidth]{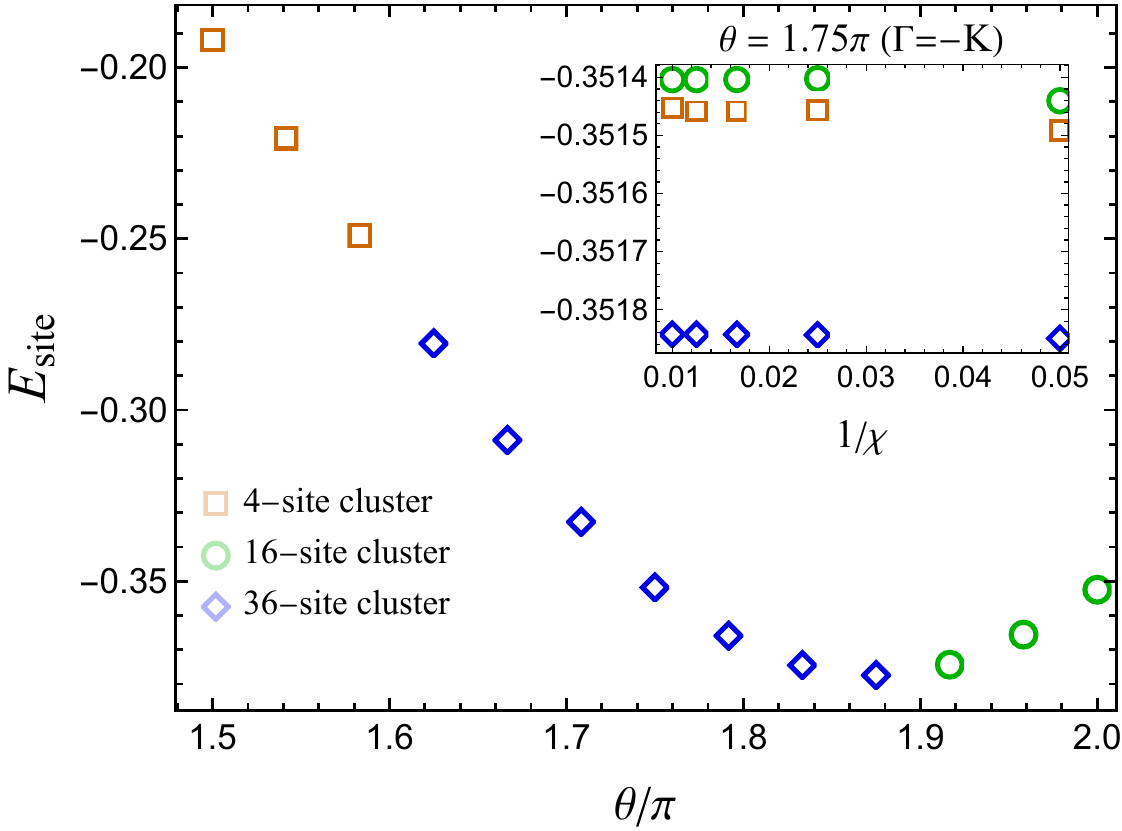}
  \caption{(Proximate) Ground state energies computed with iPEPS. In the regime with comparable Kitaev and $\Gamma$ interactions ($1.6 \pi  < \theta < 1.9\pi$), the $36$-site cluster systematically gives the lowest energies and develops a magnetic pattern similar to classical longitudinal magnetizations.
  At larger $\Gamma$ ($\theta > 1.9\pi$), the $16$-site and $4$-site clusters take over and capture a zigzag-type order. Small magnetizations are obtained near the Kitaev corner ($\theta < 1.6  \pi$) where the $4$-site cluster is preferred. The inset shows the energy convergence with the environmental bond dimension $\chi$.}
  \label{fig:ipeps}
\end{figure}

\paragraph{Proximate quantum ground states.}
The identification of the classical ground state provides a reference for understanding the quantum model.
Indeed, strong magnetic orders established in the classical limit often can persist in the quantum spin-$1/2$ case with a reduced ordering moment and shifted phase boundary.
Examples can be found in both Kitaev magnets~\cite{Osorio14, Rusnacko19, Rao21b} and highly frustrated triangular anti-ferromagnets~\cite{White07,Li18}.
\changes{Our full characterization of the classical helix tangle is especially valuable in the current problem, as its unbiased simulations are far beyond the reach of any state-of-the-art quantum algorithm.}

In particular, before $\Gamma$, hence the helicity, gets too strong, we hope iPEPS may capture the longitudinal magnetizations consistently.
We consider three tensor network ansatzes built from $4$-, $16$-, and $36$-site clusters.
These clusters cover potential competing orders such as ferromagnetic, N\'{e}el, $\sqrt{3}\times\sqrt{3}$, stripy, and zigzag type orders which are commonly found in Kitaev magnets~\cite{Rusnacko19,Rao21b,Chaloupka15}.
The $36$-site cluster can further fit two supercells of the classical helical axes.
We scan the parameter space through a simple update scheme~\cite{Jiang08,Corboz10} up to a large bond dimension $D=8$.
Typically over $600$ initializations are simulated at each $\theta$ value.
Physical quantities are measured using a CTMRG method~\cite{Nishino96,Orus09}  with environmental bond dimensions $\chi > D^2$.
For comparison, we have also examined a full update scheme~\cite{Jordan08,Osorio14} but find the improvements are limited for the symmetry-broken states.
See the SM for simulation details~\cite{SM}.

\begin{figure}[t]
  \centering
  \includegraphics[width=0.48\textwidth]{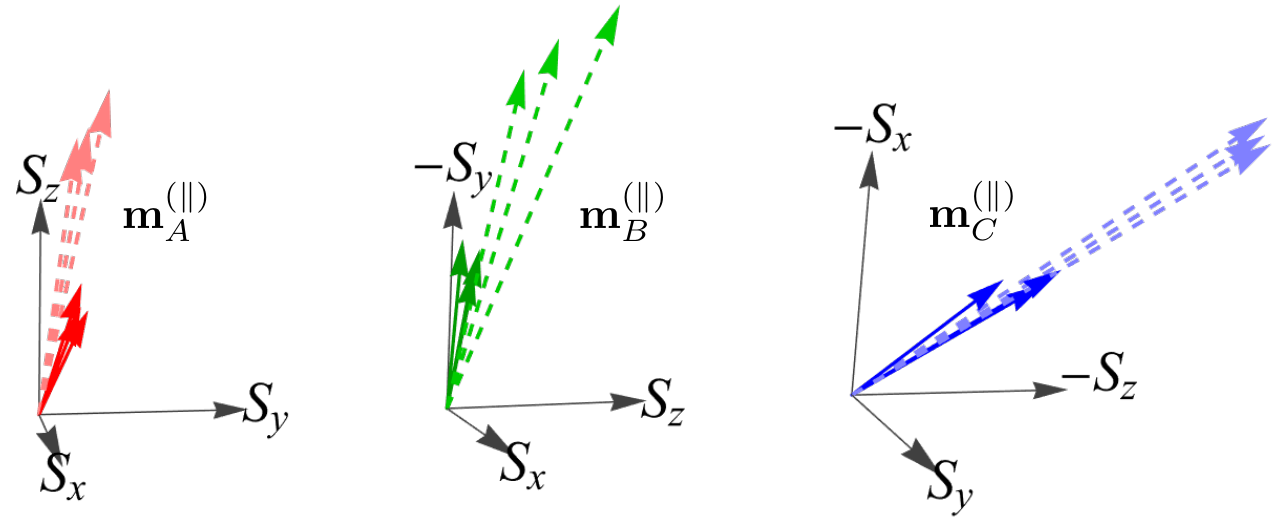}
  \caption{Comparison between quantum (solid arrows) and classical (dashed arrows) sublattice magnetizations at $\theta = 1.67\pi, 1.75\pi, 1.83\pi$ ($\Gamma = -K/\sqrt{3}, -K, -\sqrt{3}K$). 
  Orientations of the quantum magnetic moments are very close to the classical longitudinal magnetizations, while the magnitudes reduce to $0.3S \lesssim  |\mathbf{m}_{A,B,C}| \lesssim 0.5 S$ from the classical values $0.8 \lesssim  |\mathbf{m}^{\parallel}_{A,B,C}| \lesssim 1$. Axes are rotated for visualization.}
  \label{fig:compare_mags}
\end{figure}

As we show in Fig.~\ref{fig:ipeps}, within the parameter regime $1.6 \pi  < \theta < 1.9\pi$, the $36$-site cluster systematically leads to the lowest energies whose variations in $\chi$ are significantly smaller than the energy distinctions between different clusters.
This confirms the convergence of our simulations and excludes competing magnetic orders.
Remarkably, the quantum magnetic moments $\mathbf{m}_{\mu}$ display a very similar sublattice structure as the classical longitudinal magnetizations $\mathbf{m}^{\parallel}_{\mu}$~\cite{SM}. 
Their strengths reduce from the classical values $0.8 \lesssim  |\mathbf{m}^{\parallel}_{A,B,C}| \lesssim 1$ to $0.3S \lesssim |\mathbf{m}_{A,B,C}| \lesssim 0.5S$ but remain sizable to distinguish from paramagnetic states.
Orientations of the quantum magnetizations $\mathbf{m}_{\mu}$ also nearly coincide with that of the classical longitudinal moments $\mathbf{m}^{\parallel}_{\mu}$, as compared in Fig.~\ref{fig:compare_mags}.

The resemblance between the quantum and classical longitudinal magnetic moments indicates the persistence of the classical helical phase.
\changes{Although the spiral components are intractable with existing quantum algorithms,  at a weaker $\Gamma$ value $\theta \approx 1.67 \pi$, their strength remains negligible in the classical state (Fig.~\ref{fig:osc}) and affects the energy in the order of $|m^{\perp}_\mu|^2 \sim 10^{-8}$.}
Thus, one may expect the iPEPS ansatzes to remain legitimate at such $\Gamma$ values.
Moreover, given we do not observe signals of a phase transition at immediately stronger $\Gamma$ in both quantum and classical cases, the classical order can be anticipated to survive quantum fluctuations at least for a finite extent of moderate $\Gamma$ values, which is also the experimentally interested parameter regime.

In the large $\Gamma$ regime ($\theta > 1.9 \pi$), the $16$-site and $4$-site
 clusters take over with indistinguishable energies, but both show a clear zigzag type magnetization~\cite{SM}.
Nevertheless, here the iPEPS ansatzes likely fail due to growing helicity. 
The observed zigzag order may reflect a numerical artifact or a consequence of a possible quantum order-by-disorder at the $\Gamma$ point~\cite{Luo21,Discussion}.

Near the Kitaev corner ($\theta < 1.6 \pi$), we obtain small magnetizations, in line with the literature~\cite{Wang19,Gohlke20,Lee20}.

\paragraph{Summary and discussion.}
Understanding the physics of the honeycomb Kitaev-$\Gamma$ model is crucial for both interpreting experimental observations and exploring novel phases in Kitaev magnets.
In this work, we have investigated its ground-state properties through a combination of comprehensive classical and quantum simulations.
We discovered a tangle of highly structured spin double helices imprinted in the classical ground states at material-relevant $\Gamma$ values.
This helix tangle distinguished itself from typical spiral magnets~\cite{Mostovoy96,Kimura07,Khomskii09,Tokura10} by a number of prominent features, including an intricate modulation of helical axes, a spontaneous periodicity anisotropy, and a well-regulated chirality pattern.
The complete characterization of the classical ground state becomes particularly valuable given fundamental limitations in state-of-the-art quantum numerical algorithms.  
Converged iPEPS calculations reproduced magnetic moments resembling the classical longitudinal magnetization and signal persistence of the helix tangle against quantum fluctuations.
The emergence of unconventional helicity may leave fingerprints to dynamical and transport behaviors, such as spectrum broadening and splitting, long-living currents, and anomalous diffusions~\cite{Popkov17,Jepsen20,Walser12}.
  
\paragraph{Acknowledgments.}
\begin{acknowledgments}
This project is partially funded by the Deutsche Forschungsgemeinschaft (DFG, German Research Foundation) under Germany's Excellence Strategy -- EXC-2111 -- 390814868.
The research is also part of the Munich Quantum Valley, which is supported by the Bavarian state government with funds from the Hightech Agenda Bayern Plus.
N.R., L.P., and K.L. also acknowledge support from FP7/ERC Consolidator Grant QSIMCORR, No. 771891.
Our numerical simulations make use of the QSpace tensor library~\cite{Weichselbaum12, Weichselbaum20}, the TKSVM library~\cite{Greitemann19,Liu19}, and the ALPSCore library~\cite{Gaenko17}.
The simulations were performed on the KCS cluster at Leibniz-Rechenzentrum (LRZ) and the ASC cluster at Arnold Sommerfeld Center.
The data used in this work will be made available~\cite{Data_repo2023}.
\end{acknowledgments}

\bibliographystyle{apsrev4-1}
\bibliography{jkgm.bib}

\begin{thebibliography}{73}
\bibitem [{Dis()}]{Discussion}%
  \BibitemOpen
  \href@noop {} {}\bibinfo {note} {Simulations at the $\Gamma$ point are prone
  to boundary conditions and algorithms. Private communications with Hong-Hao
  Tu and Ji-Ze Zhao.}\BibitemShut {Stop}%
\bibitem [{Map()}]{Mapping}%
  \BibitemOpen
  \href@noop {} {}\bibinfo {note} {This is a similar to the mapping between the
  ferromagnetic and anti-ferromagnetic Heisenberg model on a bipartite
  lattice.}\BibitemShut {Stop}%
\end{thebibliography}

%
%
%
\onecolumngrid
\clearpage
\makeatletter
\begin{center}
  \textbf{\large --- Supplementary Materials ---\\[0.5em]Tangle of Spin Double Helices in the Honeycomb Kitaev-$\Gamma$ Model}\\[1em]

Jheng-Wei Li$^{1,2}$, Nihal Rao$^{1,2}$, Jan von Delft$^{1,2}$, Lode Pollet$^{1,2,3}$, and Ke Liu$^{1,2}$

\vspace{5pt}

\scalebox{0.95}{\centering \emph{$^1$Arnold Sommerfeld Center for Theoretical Physics,  University of Munich, Theresienstr. 37, 80333 M\"{u}nchen, Germany}}

\scalebox{0.95}{\centering \emph{$^2$Munich Center for Quantum Science and Technology (MCQST), Schellingstr. 4, 80799 M\"{u}nchen, Germany}}

\scalebox{0.95}{\centering \emph{$^3$Wilczek Quantum Center, School of Physics and Astronomy, Shanghai Jiao Tong University, Shanghai 200240, China}}

  \thispagestyle{titlepage}
\end{center}
\setcounter{equation}{0}
\setcounter{figure}{0}
\setcounter{table}{0}
\setcounter{page}{1}
\setcounter{section}{0}
\renewcommand{\theequation}{S\arabic{equation}}
\renewcommand{\thefigure}{S\arabic{figure}}
\renewcommand{\thetable}{S\arabic{table}}
\renewcommand{\thesection}{S.\Roman{section}}

\section{Classical ground states}
\subsection{Details of the classical simulations}
Simulating highly frustrated models with large system sizes is in general a challenging task.
To ensure we access the correct classical ground states, we first use a parallel tempering Monte Carlo (PTMC) method to generate spin configurations at a low temperature $T = 10^{-5} \sqrt{K^2 + \Gamma^2}$ and then cool the system to $T \rightarrow 0$ by eliminating remnant small thermal noises (see below).
We mostly consider lattices with linear spacing $L = 72$ ($10, 386$ spins) on a torus, and have also checked larger lattices up to $L = 108$ ($23,328$ spins) at particular parameter points.
Such large systems sizes and low temperatures are crucial to manifest the spin helices.

We use parallel tempering jointly with heat bath and over-relaxation algorithms to equilibrate the system.
$N_T = 256$ logarithmically equidistant temperatures are used to ensure efficient iterations between different temperatures~\cite{Katzgraber06}.
Typically $10^7$ Monte Carlo sweeps are performed in an individual run.
All independent runs have converged to the same states, confirming the ergodicity of our simulations.

The PTMC generates spin configurations lying slightly above the classical ground states by an energy scale $\Delta E \sim 10^{-5}$ preset by the lowest simulated temperature.
We cool the system to further approach the ground states by iteratively aligning spins along their local molecular fields $\mathbf{B}^{\rm loc}_{i}$~\cite{Janssen16},
\begin{align}
    \mathbf{S}^{\rm new}_{i} = \frac{\mathbf{B}^{\rm loc}_{i}}{|\mathbf{B}^{\rm loc}_{i}|}|\mathbf{S}^{\rm old}_{i}|.
\end{align}
Here $\mathbf{B}^{\rm loc}_{i} = \sum_{\langle i j\rangle_{\gamma}} \hat{J}^{\gamma}_{ij}\mathbf{S}_{j}$ and the exchange tensors $\hat{J}^\gamma_{ij}$ are reproduced for convenience 
\begin{align}
\hat{J}^x_{ij} = \left[ 
\begin{smallmatrix}
K & &  \\
 &  & \Gamma \\
 & \Gamma  & 
\end{smallmatrix}
\right], \ 
\hat{J}^y_{ij} = \left[ 
\begin{smallmatrix}
 & & \Gamma \\
 & K &  \\
 \Gamma &  & 
\end{smallmatrix}
\right],  
\hat{J}^z_{ij} = \left[ 
\begin{smallmatrix}
  & \Gamma &  \\
\Gamma & &  \\
 &  & K
\end{smallmatrix}
\right].
\end{align}

\begin{figure*}[!ht]
    \centering
    \includegraphics[width=0.9\textwidth]{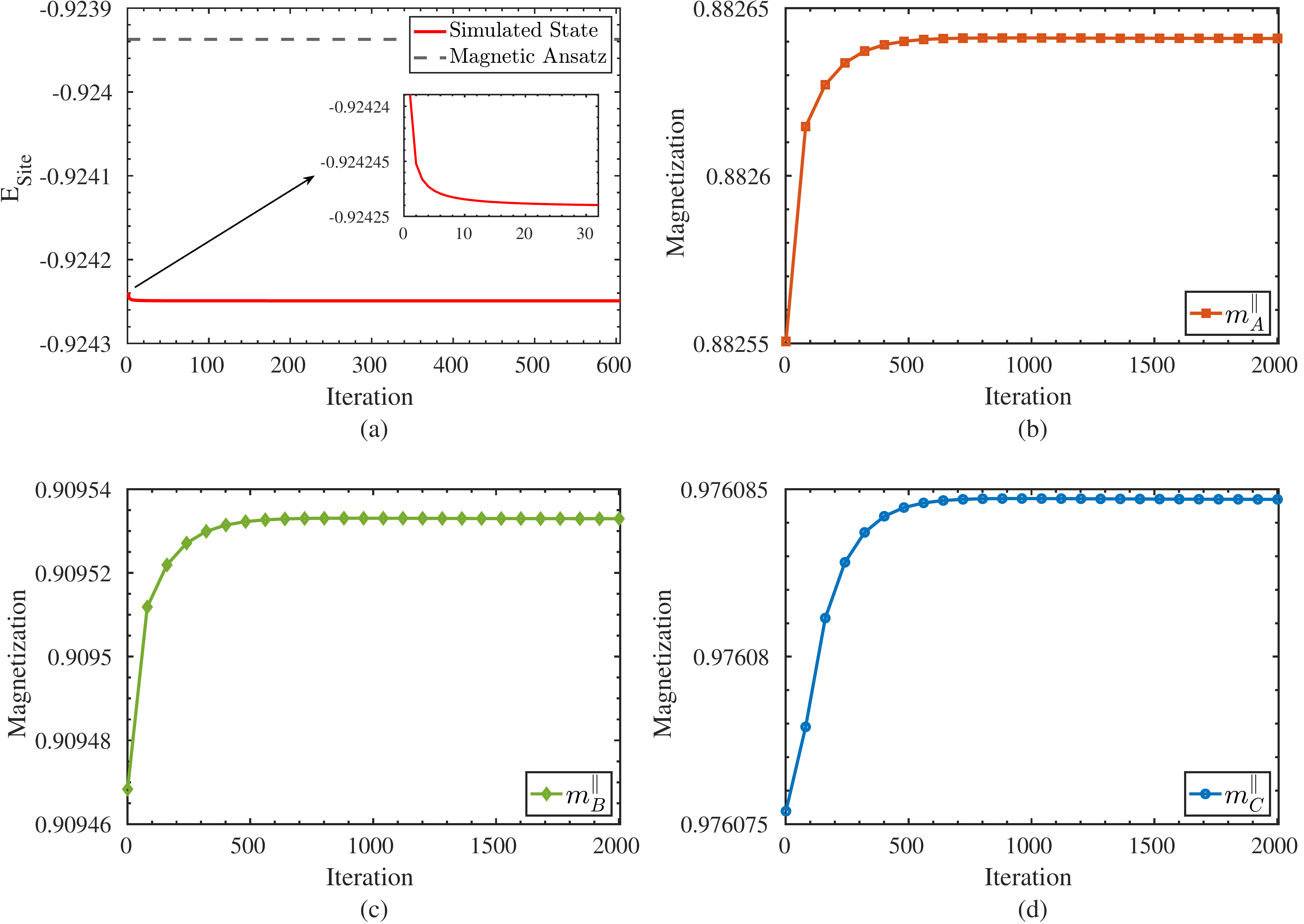}
   \caption{Cooling of a classical state simulated on a $L=72$ lattice at $\Gamma = -K$ ($\theta = 1.75 \pi$), with $|\mathbf{S}_i|=1$. (a) Convergence of the energy. The energy per spin is $E_{\rm site} = -0.92423895$ before the cooling and is $E_{\rm site} = -0.92424917$ after the cooling (solid line); both are lower than the ansatz energy $E_{\rm site} =  -0.92393734$ (dashed line). 
   The inset magnifies the evolution in a short time period.
   (b-d) Convergence of the sublattice (longitudinal) magnetizations, $m^{\parallel}_{A,B,C}$, of the helical axes. These magnetizations converge to values below unity owing to non-vanishing spiral magnetizations $m^{\perp}_{A,B,C}$.}
       \label{fig:cooling}
\end{figure*}

We continue to cool the spins until the maximum difference in energies of the spin configurations between successive iterations $|E^{\rm old} - E^{\rm new}|_{\rm max}$ is less than $10^{-14}$.
Evolutions of energies and magnetizations during the cooling are shown in Fig.~\ref{fig:cooling} for example.
At each fixed parameter point, we examine a number of statistically uncorrelated configurations and find identical energies up to the numerical precision, which reaffirms the ergodicity of our simulations.

\begin{figure*}[b]
  \centering
  \includegraphics[width=0.9\textwidth]{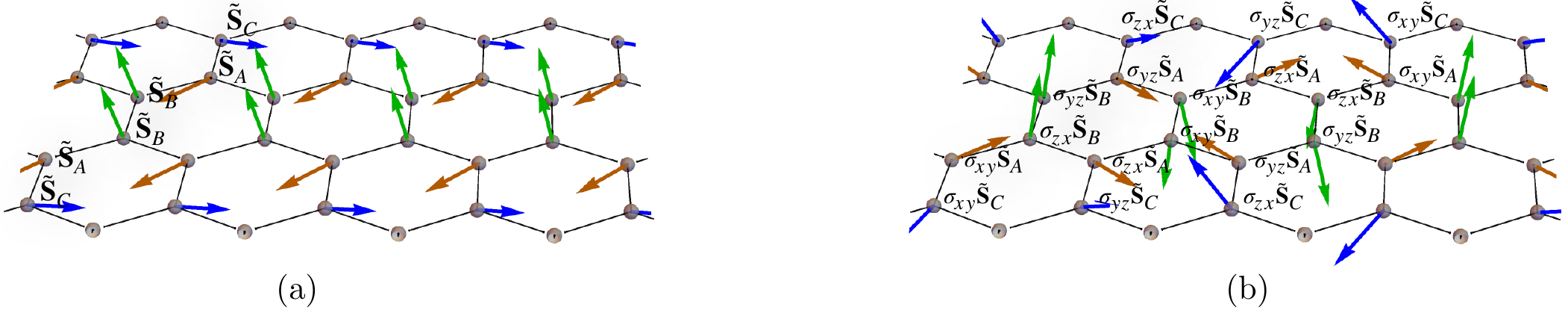}
  \caption{Two degenerate magnetic ansatzes obtained by enforcing a $3 \times 3$ periodic cluster at $K = -\Gamma$. (a) The three ansatzs spins $\tilde{\mathbf{S}}_A, \tilde{\mathbf{S}}_B, \tilde{\mathbf{S}}_C$ form a $6$-site repeating pattern $C{\text -}A{\text -}B{\text -}B{\text -}A{\text -}C$. (b) An $18$-site degenerate repeating pattern is obtained by applying the hidden symmetry.}
  \label{fig:ansatzes}
\end{figure*}

\begin{figure*}[t]
  \centering
  \includegraphics[width=0.45\textwidth]{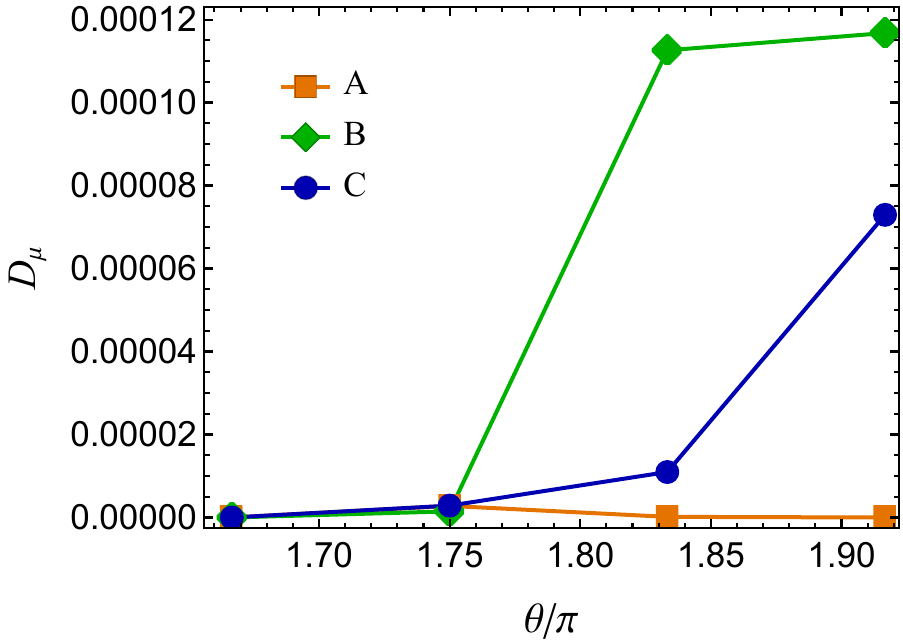}
  \caption{Cosine de-similarity $D_\mu \in [0, 1]$ between the ansatz spin orientations $\tilde{\mathbf{S}}_{A,B,C}$ and the sublattice magnetizations $\mathbf{m}^{\parallel}_{A,B,C}$ simulated on a $L = 72$ lattice. 
  $D_\mu$ generally rises with increasing $\Gamma$, which is expected due to the stronger spiral magnitudes.
  Nevertheless, even at the large $\Gamma$ value $\theta \approx 1.92 \pi$, the de-similarities remain remarkably small, and the magnetic ansatzes can still provide a proximate description of the correct sublattice structure.}
  \label{fig:deSimilarity}
\end{figure*}

\subsection{Simulated states vs. translational invariant ansatzes}
A common strategy for finding classical magnetic ground states is to minimize the Hamiltonian with small periodic clusters, while the choice of those clusters can be guided by small-size trial simulations.
This approach fails to capture the true ground states of the Kitaev-$\Gamma$ model due to the emergent helicity.
Nonetheless, it still leads to useful information about longitudinal magnetization of the helical phase, in particular the sublattice structure and orientations of the helical axes.

By \emph{enforcing} a $3 \times 3$ cluster, one obtains two translational invariant ansatzes related by the hidden symmetry
$\sigma^{\alpha\beta}_i \hat{J}^{\gamma}_{ij} \sigma^{\alpha\beta}_j = \hat{J}^{\gamma}_{ij}$ and 
$\sigma^{\alpha\gamma}_i \hat{J}^{\gamma}_{ij} \sigma^{\gamma\beta}_j = \hat{J}^{\gamma}_{ij}$,
as shown in Fig.~\ref{fig:ansatzes}.
These two states have a $6$-site and an $18$-site repeating pattern, respectively, and can be represented via three elementary orientations $\tilde{\mathbf{S}}_A, \tilde{\mathbf{S}}_B, \tilde{\mathbf{S}}_C$, where the narrow tilde symbols are used to distinguish from spins $\mathbf{S}$ in actual simulations.

Express $\tilde{S}_{A,B,C}$ with angle parameters as
$\tilde{S}_{A,B,C} = \left(\sin\alpha\sin\beta, \, \sin\alpha\cos\beta, \, \cos\alpha \right)_{A,B,C}$.
The ansatz states can be solved by minimizing
\begin{align}
	\tilde{E}_{\rm site} = 
	& \frac{K}{6} \bigg[2 \cos\alpha _A \cos\alpha_B + 2 \sin\alpha_A \big(\sin\beta_A \sin\alpha_B \sin\beta_B + \cos\beta_A \sin\alpha_C \cos\beta_C \big) + \sin^2\alpha_B \cos^2\beta_B +\sin ^2\alpha_C \sin^2\beta_C + \cos^2\alpha_C\bigg]
	\nonumber \\
	& + \frac{\Gamma}{3} \bigg[\sin\alpha_B \big(\sin\alpha_A \sin\left(\beta_A+\beta_B\right)+\cos\alpha_A \cos\beta_B \big) 
	  + \cos\alpha_B \big(\sin\alpha_A \cos\beta_A + \sin\alpha_B \sin\beta_B \big)
	  + \cos\alpha_C \big(\sin\alpha_A \sin\beta_A 
	 \nonumber \\
	 & \qquad + \sin\alpha_C \cos\beta_C\big)
	  + \sin\alpha_C \sin\beta_C \big(\cos\alpha_A + \sin\alpha_C \cos\beta_C\big)\bigg].
\end{align}
The solutions of $\tilde{S}_{A,B,C}$ vary against the ratio of $K/\Gamma$, and their orientations are non-collinear and non-coplanar.

Both the limitation and the value of the two magnetic ansatzes in Fig.~\ref{fig:ansatzes} need to be emphasized.
On the one hand, their energy is very close to the ground-state energy.
For instance, at $K=-\Gamma$ and on a $L=72$ lattice, $\min{\tilde{E}_{\rm site}}$ is only higher than the simulated energy by an amount of $\sim 10^{-4}$, as compared in Fig.~\ref{fig:cooling}(a).
However, this does {\it not} mean these translational invariant states could represent the true ground states.
Instead, they only approximate the longitudinal components of the ground-state ordering but completely elude the transverse sector. 

On the other hand, these ansatzes approximate the longitudinal magnetic moments
$\mathbf{m}^{\parallel}_\mu = \frac{1}{N_{\rm cell}} \sum_{\rm cells} \mathbf{S}_\mu =  \mathbf{S}^{\parallel}_\mu$,
which defines the helical axes, with very high precision.
The ansatz spins $\tilde{S}_{A,B,C}$ are nearly parallel with the longitudinal magnetization of actual spins.
This is confirmed by introducing a quantitive measure 
\begin{align}\label{eq:D_sim}
	D_{\mu} = \frac{1}{2}\left(1- \tilde{\mathbf{S}}_\mu\cdot \frac{\mathbf{m}^{\parallel}_{\mu}}{ \lvert\mathbf{m}^{\parallel}_{\mu}\rvert} \right) \in \left[0,1\right], \
\end{align}
which is a cosine de-similarity between the {\it orientations} of $\tilde{\mathbf{S}}_\mu$ and $\mathbf{m}^{\parallel}_\mu$.
Note that $\mathbf{m}^{\parallel}_\mu$ is renormalized in Eq.~\eqref{eq:D_sim}, by which the difference in magnitude has been ignored.

As measured in Fig.~\ref{fig:deSimilarity}, even in the region where the strength of the spiral magnetization $\mathbf{m}^{\perp}_\mu$ has become comparable with that of the longitudinal $\mathbf{m}^{\parallel}_\mu$ (see Fig.~2 in the main text), $D_{\mu}$ remains extremely small.
This hence provides an intuitive understanding of the classical ground states: 
The longitudinal order of spins can be estimated using the magnetic ansatzes in Fig.~\ref{fig:ansatzes}, while actual spins form helices and swirl about the longitudinal moments according to Eq.~(5) in the main text.

The source of this helicity may be understood by the exchange frustrations between $\Gamma$ matrices. 
For example, on a $z$ bond, the $\Gamma$ interaction can be rewritten as
$\Gamma_z = \left(
\begin{smallmatrix}
 0 & 1 & 0\\
 1 & 0 & 0\\
 0 & 0 & 0
\end{smallmatrix}
\right) 
= -L_z \sigma_{xz}
$,
where 
$L_z = \left(
\begin{smallmatrix}
 0 & -1 & 0\\
 1 & 0 & 0\\
 0 & 0 & 0
\end{smallmatrix}
\right) 
$
is a generator of ${\rm SO}(3)$ rotations about the $S_z$ axis, and
$\sigma_{xz} = \left(
\begin{smallmatrix}
 1 & 0 & 0\\
 0 & -1 & 0\\
 0 & 0 & 0
\end{smallmatrix}
\right) 
$
is a mirror respect to the spin $xy$-plane and brings ${\rm SO}(3)$ rotations into ${\rm O}(3)$ rotations. 
Thus, a spiral structure may be intrinsically favored by $\Gamma$ interactions.
Consistently, the spiral magnetization $m^\perp_\mu$ increases with $\Gamma$.
Furthermore, as the three types of lattice bonds host different $\Gamma$ matrices, there is competition between different rotation axes.
An interplay between helicity and competing interactions may have led to the modulation in the $18$ helical axes.
Nevertheless, a finite competing $K$ is still necessary.
If $K$ vanishes, the Hamiltonian reduces to a pure $\Gamma$ model featuring an emergent local $Z_2$ symmetry.
All magnetic orders will be destroyed due to extensive degenerate ground states, and the system becomes a classical $\Gamma$ spin liquid.

\begin{figure*}[t]
  \centering
  \includegraphics[width=0.95\textwidth]{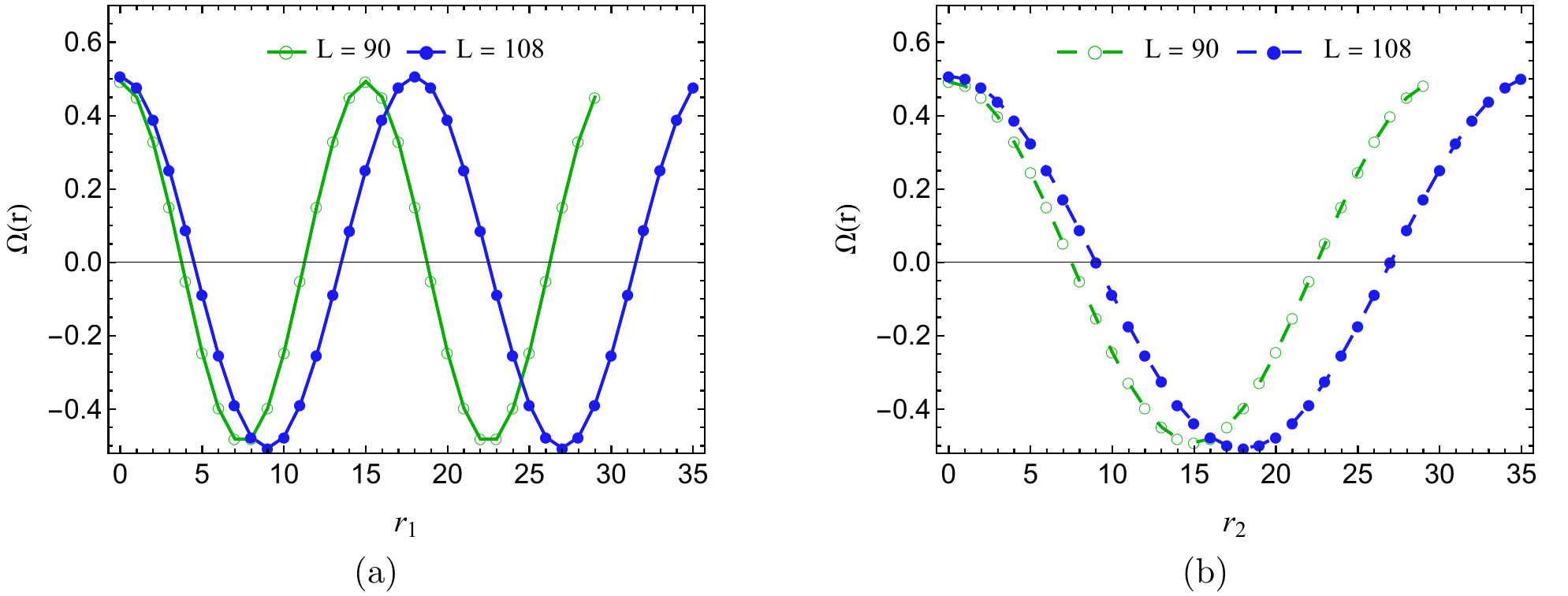}
  \caption{\changes{Finite-size dependence and spontaneous anisotropy of the helix pitches. The spiral correlation function $\Omega_\mu({\bf r})$ for type-B helices is measured at $K=-\Gamma$ along the direction of two independent lattice vectors. (a) The oscillations show two periods along a spontaneously chosen $r_1$ direction, confirming the pitch sizes of $\frac{L}{6} = 15, \, 18$ supercells ($45$ and $54$ lattice spacings) for lattices $L = 90, \, 108$, respectively. 
  (b) There is only a single period along the $r_2$ direction, with pitch sizes of $\frac{L}{3} = 30, \, 36$ supercells ($90$ and $108$ lattice spacings), justifying a spontaneous anisotropy in periodicity. Type-A and type-C helices have the same periodicity but differ in the oscillation amplitude as $|\Omega_\mu|\propto m^{\perp}_\mu$; $r_1, \, r_2 = 0, 1, \dots, \frac{L}{3} -1$ due to the $3\times3$ supercell structure.}}
  \label{fig:SM_osc}
\end{figure*}

\subsection{Genuine incommensuration}

This unusual helical phase is genuinely incommensurate. The transverse components of spins are captured by a spiral ansatz (reproduced from Eq.~(5) in the main text for convenience)
\begin{align}\label{eq:SM_S_trans}
	\widetilde{\mathbf S}^{\perp}_{\mu,\mathbf{r}} = m^{\perp}_\mu 
	\left(\cos\tfrac{6\pi}{L}(2r_1 + r_2), \, 
	 {\rm sgn}(\Gamma)^\eta \sin\tfrac{6\pi}{L}(2r_1 + r_2), \, 0
	\right),
\end{align} 
that can be verified explicitly.
According to this ansatz, the helix pitch sizes are $\frac{L}{6}$ and $\frac{L}{3}$ supercells along the directions of the two lattice vectors, respectively.
Namely, the helix pitches are spontaneously anisotropic and, in any finite-size simulation, shall grow with the lattice.
In the main text, we have shown results for a $L=72$ lattice, where the pitch sizes are $12$ and $24$ supercells. 
Here, we provide additional simulations for lattices $L = 90, \, 108$ and again measure the spiral correlation function
$\Omega_{\mu} (\mathbf{r}) = \frac{1}{m^{\perp}_\mu} \left( \mathbf{S}_{\mu, \mathbf{r}_0} \cdot \mathbf{S}_{\mu, \mathbf{r}_0 + \mathbf{r}} - |\mathbf{m}^{\parallel}_\mu|^2\right)$.
As seen from Fig.~\ref{fig:SM_osc}, the helices complete two periods along a spontaneously determined $r_1$ direction, leading to helix pitches of $\frac{L}{6} = 15$ and $18$ supercells ($45$ and $54$ lattice spacings) for the $L = 90$ and $L=108$ lattices, respectively.
Consistently, the helices complete one period with a single pitch across the entire lattice ($30$ and $36$ supercells) long the $r_2$ direction.
These results unambiguously justify the emergent helicity and our ansatz Eq.~\eqref{eq:SM_S_trans}.

Note that although the magnitude ($m^{\perp}_\mu$) of the spiral components depends on the ratio of $K/\Gamma$ as we measured in Fig.~2 in the main text, the pitch sizes do not. 
The ansatz Eq.~\eqref{eq:SM_S_trans} is valid in the entire helical phase.

\section{Proximate quantum ground states}

\begin{figure*}
  \centering
  \includegraphics[width=1\textwidth]{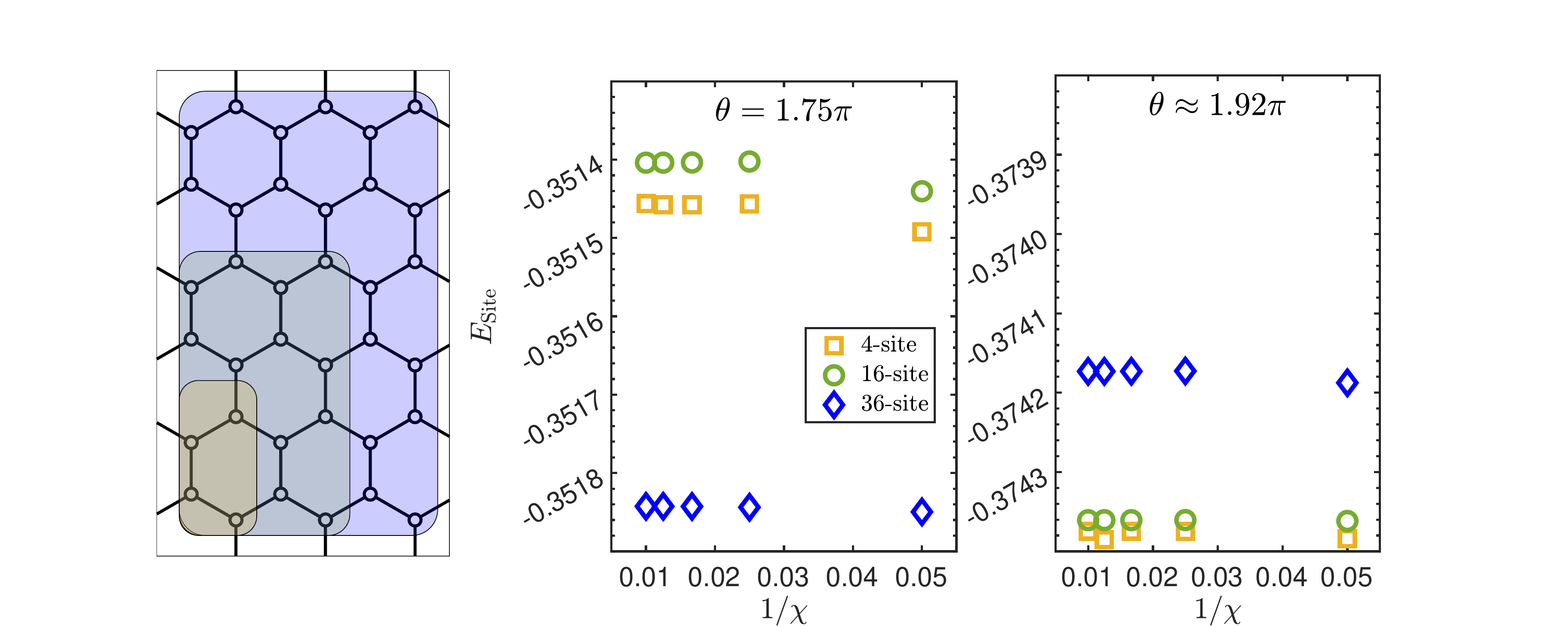}
  \caption{Geometries of the iPEPS ansatzes and energy convergence. The three clusters are indicated by shadings in the left panel. The $36$-site cluster leads to the lowest energy at the moderate $\Gamma = -K$ ($\theta = 1.75 \pi$), but is superseded by the $16$-site and $4$-site clusters at a stronger $\Gamma \approx - 3.73K$ ($\theta \approx 1.92 \pi$).}
  \label{fig:clusters}
\end{figure*}

\subsection{Details of iPEPS simulations}
Our iPEPS ansatzes are formulated on a brick-wall lattice which maps onto the honeycomb lattice by introducing a trivial index on each tensor~\cite{Corboz12, Osorio14}.
Three tensor ansatzes with a $4$-site, a $16$-site, and a $36$-site geometry are considered, as illustrated in Fig.~\ref{fig:clusters} (left panel).
The former two geometries are chosen to probe ferromagnetic, N\'{e}el, stripy, and zigzag-type orders, while the latter can (additionally) capture $\sqrt{3}\times\sqrt{3}$-type orders as well as the longitudinal sublattice magnetization of the helical axes.
We initialize our simulations using both random tensors and the corresponding classical ground states. Typically over $600$ initializations are examined for each $\theta$ value.

We adopt the simple update scheme~\cite{Jiang08,Corboz10} to scan the parameter space and run the simulations with up to a bond dimension $D=8$.
In addition, we have also compared the results with the full update scheme~\cite{Jordan08,Osorio14} at selected parameter points.
Although the full update method can noticeably improve simulations at the Kitaev spin liquid points~\cite{Osorio14}, we nevertheless find that the improvements are limited for the present symmetry-broken states.

We use the corner transfer matrix renormalization group (CTMRG) method~\cite{Nishino96,Orus09} to determine the value of physical quantities.
Large environmental dimensions $\chi > D^2$ are examined for ensuring the CTMRG convergence as exemplified in Fig.~\ref{fig:clusters} for a moderate and a large $\Gamma$ value.
In the moderate $\Gamma$ regime, $1.6 \pi  < \theta < 1.9\pi$ the $36$-site cluster provides the best approximation to the ground state and can be clearly distinguished from the other two clusters.
The $16$-site and $4$-site clusters lead to the lowest energies for the large $\Gamma$ regime $\theta > 1.9 \pi$ with nearly degenerate values.
Their energy difference is comparable to the energy variation due to the finite $\chi$ approximation, while convergences of the $16$-site cluster appear to be better over this regime.
Nevertheless, there we expect the iPEPS to fail because of growing helicity. Hence the change from the $36$-site ansatz to the $16$-site or $4$-site one does not necessarily reflect a phase transition.

\begin{figure*}[b]
  \centering
  \includegraphics[width=1\textwidth]{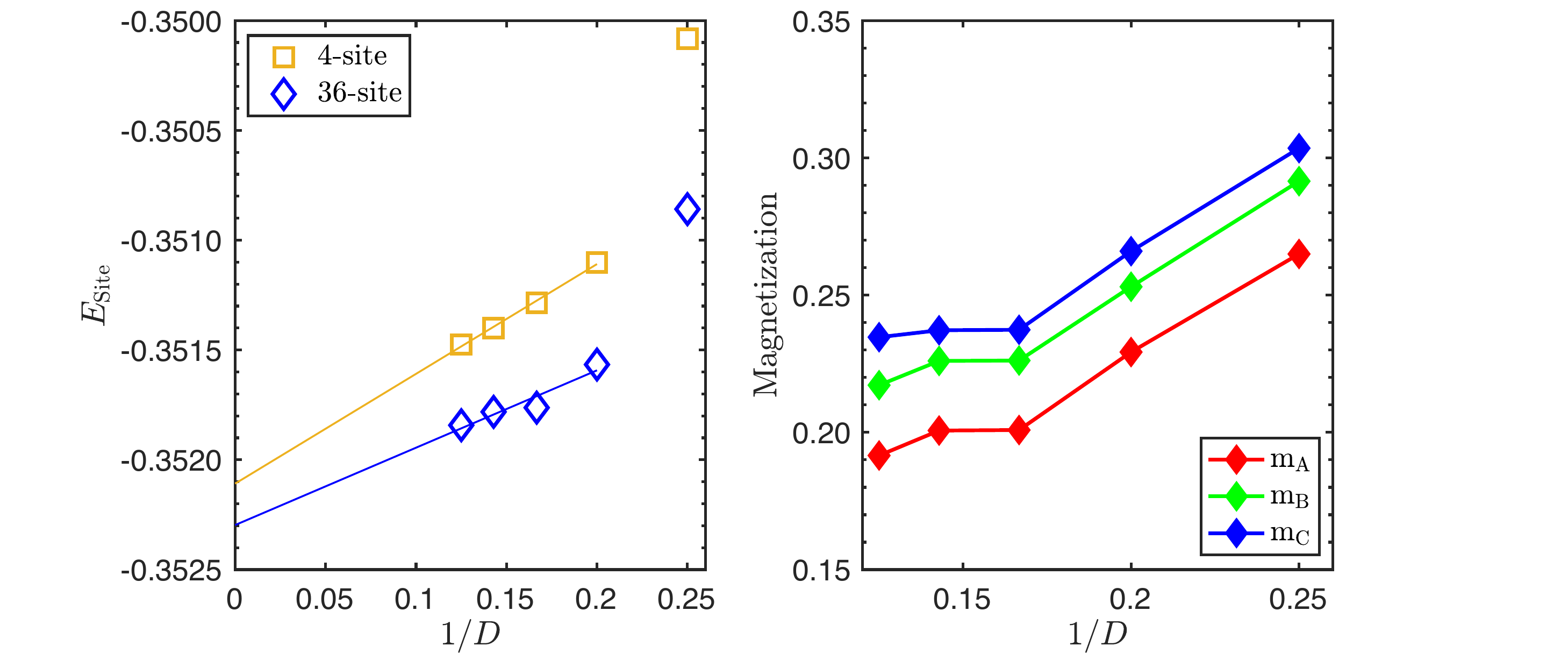}
  \caption{\changes{Dependence of the ground-state energy ($E_{\rm site}$) and sublattice magnetizations ($m_{A,B,C}$) on the bond dimension ($D$), at $\Gamma = -K$. Left panel: Energies of the $4$-site and $36$-site clusters are compared for $D = 4,5,6,7,8$ with a trend of convergence. The latter robustly leads to lower energy at all $D$s. Linear extrapolations in $1/D$ towards $D\rightarrow \infty$ indicate the lower bounds of $E_{\rm site}$.
  	 Right panel: Sublattice magnetizations of the $36$-site cluster become stable from $D = 6$. Aside from the orientations (Fig.~5), their magnitudes $m_A < m_B < m_C$ also reproduce the relation of the classical longitudinal magnetic moments.}}
  \label{fig:D_dependence}
\end{figure*}

Fig.~\ref{fig:D_dependence} shows the $D$-dependence of the ground-state energy and sublattice magnetizations at the parameter point $\Gamma = -K$ where the classical helicity becomes noticeable but remains mild.
Energies of the $4$-site and the $36$-site clusters are compared for $D=4, 5, 6, 7, 8$, while results of the $16$-site one are always quasi-degenerate with the former.
One can see clearly that the large $36$-site cluster robustly leads to lower variational energy and shows a trend of convergence.
Moreover, its associated sublattice magnetizations become stable at larger $D$, and their magnitudes, with $m_A < m_B < m_C$ as the longitudinal magnetizations in the classical helical phase.

\begin{figure}[t]
\centering
  \includegraphics[width=0.75\textwidth]{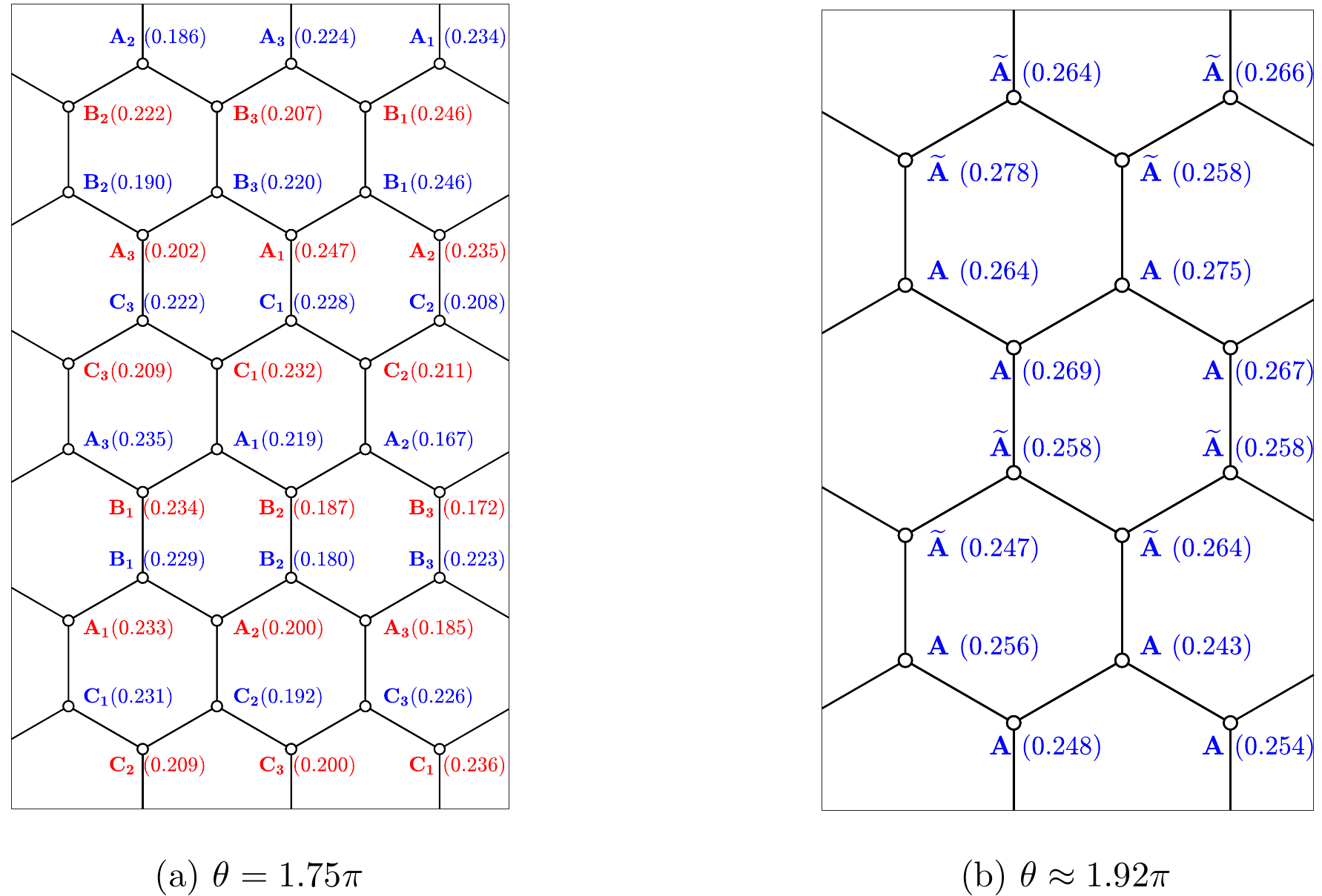}
\caption{Magnetic structures computed by the iPEPS at $\theta = 1.75\pi, 1.92\pi$ ($\Gamma = -K, -3.73K$), with $S = \frac{1}{2}$.
(a) Sites in the $36$-site cluster are labeled similarly to the $3\times 3$ unit cell of the classical helical axes. $A_j, B_j, C_j$ indicate three elementary orientations of the magnetic moments. The numbers show the strength of the corresponding sublattice magnetization.
(b) A zigzag order captured by the $16$-site cluster. $A$ and $\tilde{A}$ denote two opposite magnetic moments.}
\label{fig:ipeps_mag}
\end{figure}

\subsection{Comparison of quantum and classical magnetic moments}

Very large system sizes are required to represent a helix state, while the situation in the Kitaev-$\Gamma$ model is especially challenging.
In our classical simulations on the $L = 72$ lattice, each helix is ``only'' formed by $24$ spins due to the $3\times 3$ modulation of the helical axes.
Faithful simulations of the spin-$1/2$ Kitaev-$\Gamma$ model are hence far beyond the capabilities of available quantum algorithms.

Nevertheless, our detailed characterization of the classical helicity provides a possibility to gain insight into the quantum ground states in regimes where the helicity remains mild in strength.
As we measured in the main text by the oscillation parameter $\Omega_\mu$ (Fig.~3), the classical spiral magnetization is about $m^{\perp}_\mu \sim 10^{-4}$ at $\Gamma = -K/\sqrt{3}$ ($\theta \approx 1.67\pi$).
Since this quantity affects the energy of a two-body Hamiltonian in quadratic form, i.e., in a magnitude $\mathcal{O}(10^{-8})$, we expect the iPEPS ansatzes to remain legitimate at such a $\Gamma$ value despite that the ground state is not translationally invariant.
Provided the simulations do not detect signals of a phase transition when slightly increasing $\Gamma$, one naturally expects the same physics to manifest for a finite parameter regime. 
Then by comparing the structures of the classical and quantum moments, we may infer whether the classical order is immediately destroyed by quantum fluctuations.

The magnetic pattern captured by our converged simulations using the $36$-site cluster shows remarkable resemblances to the longitudinal magnetization of the classical helical axes.
In addition to the orientations of its magnetic moments presented in the main text (Fig.~5.), we show in Fig.~\ref{fig:ipeps_mag} the structure and magnitudes of its sublattice magnetizations.
The $16$-site and $4$-site clusters capture a zigzag-type order in the large $\Gamma$ region, which is also depicted in Fig.~\ref{fig:ipeps_mag}.


\end{document}